\documentclass[a4paper, 11pt]{article}
\usepackage{cite}
\usepackage{pos}
\usepackage{physics}
\usepackage{url}
\usepackage[frozencache,cachedir=minted-files]{minted}
\usepackage{subcaption}

\usepackage{color}
\usepackage{tikz}
\usetikzlibrary{decorations.pathmorphing,decorations.markings}
\usetikzlibrary{decorations.pathreplacing}
\usetikzlibrary{positioning, calc, shapes}
\usetikzlibrary{shapes.geometric, arrows}
    \tikzstyle{startstop} = [rectangle, rounded corners, minimum width=2cm,text centered, draw=black, fill=red!30]
    \tikzstyle{roundbox} = [rectangle, rounded corners, minimum height=0.6cm, minimum width=5cm,text centered, draw=black, fill=blue!30]

    \tikzstyle{check} = [rectangle, minimum width=2.0cm, text centered, draw=black, fill=blue!30]
    \tikzstyle{iout} = [trapezium, trapezium left angle=-70, trapezium right angle=-110, minimum width=3cm, text centered, draw=black, fill=blue!30]
    \tikzstyle{io} = [trapezium, trapezium left angle=70, trapezium right angle=110, minimum width=3cm, minimum height=1cm, text centered, draw=black, fill=blue!30]
    \tikzstyle{nnpdf} = [rectangle, minimum width=3cm, minimum height=1cm, text centered, draw=black, fill=orange!30]
    \tikzstyle{n3py} = [rectangle, minimum width=3cm, minimum height=1cm, text centered, draw=black, fill=green!30]
    \tikzstyle{n3cpp} = [rectangle, minimum width=3cm, minimum height=1cm, text centered, draw=black, fill=blue!30]
    \tikzstyle{procblue} = [rectangle, minimum width=3cm, minimum height=1cm, text centered, draw=black, fill=blue!30]
    \tikzstyle{fitted} = [rectangle, minimum width=5cm, minimum height=1cm, text centered, draw=black, fill=red!30]
    \tikzstyle{fixed} = [rectangle, minimum width=5cm, minimum height=1cm, text centered, draw=black, fill=blue!30]
    \tikzstyle{arrow} = [thick,->,>=stealth]

\definecolor{darkgreen}{rgb}{0.0, 0.5, 0.13}
\definecolor{darkred}{rgb}{0.55, 0.0, 0.0}
\title{VegasFlow: accelerating Monte Carlo simulation across platforms}
\ShortTitle{VegasFlow}

\author*[a]{Juan M. Cruz-Martinez}
\author[a]{Stefano Carrazza}

\affiliation[a]{TIF Lab, Dipartimento di Fisica, Universit\`a degli Studi di Milano and INFN Sezione di Milano \\ Via
Celoria 16, 20133, Milano, Italy}

\emailAdd{juan.cruz@mi.infn.it}
\emailAdd{stefano.carrazza@mi.infn.it}

\abstract{In this work we demonstrate the usage of the VegasFlow library on multidevice situations: multi-GPU in
    one single node and multi-node in a cluster.
    VegasFlow is a new software for the fast evaluation of highly parallelizable integrals based on Monte Carlo integration.
    It is inspired by the Vegas algorithm, very often used as the driver of cross section integrations, and based on
    Google's powerful TensorFlow library.
    In this proceedings we consider a typical multi-GPU configuration to benchmark how different batch sizes can
    increase (or decrease) the performance on a Leading Order example integration.
}

\FullConference{%
  40th International Conference on High Energy physics - ICHEP2020\\
  July 28 - August 6, 2020\\
  Prague, Czech Republic (virtual meeting)
}

\begin{document}

\maketitle

\section{Introduction}
State-of-the-art computations in High Energy Physics (HEP) require computing complex multi-dimensional
integrals numerically, as the analytical result is often not known.
Monte Carlo (MC) algorithms are generally the option of choice for these kind of integrals, be it when considering
HEP applications or elsewhere,
as the error of such algorithms does not grow with the number of dimensions.

In particular, in the HEP literature, MC methods based on the idea of importance sampling are widespread
as they combine the robustness of MC algorithms for high dimensional situations with the flexibility of
adaptive grids.

The Vegas algorithm~\cite{Lepage:1977sw, Lepage:1980dq}
is one of the main drivers for multi-purpose parton level event generation programs based on fixed order calculations
such as MCFM~\cite{Campbell:2015qma, Campbell:2019dru} or NNLOJET~\cite{Gehrmann:2018szu}
and is also present in more general tools such as MG5\_aMC@NLO~\cite{Alwall:2014hca} and Sherpa~\cite{Gleisberg:2008ta}.
Whereas the original implementation of the algorithm was written for a single CPU, nowadays it is usually
implemented to take advantage of multi-threading CPUs and distributed computing.
Indeed, MC computations are what is informally known as ``embarrassingly parallel''.

However, the parallelization of a computation over multiple CPUs does not decrease the number of CPU-hours
required to complete a computation and the cost of such calculations is driving the budget of
big science experiments such as ATLAS or CMS~\cite{Buckley:2019wov}.
With VegasFlow~\cite{Carrazza:2020rdn,juan_cruz_martinez_2020_3691927} we implement
the importance sampling techniques from Vegas to run both in CPUs and GPUs, enabling further acceleration of complicated integrals
by leveraging the abstraction possibilities offered by TensorFlow~\cite{tensorflow2015:whitepaper}.

\section{A toolset for a new generation of Monte Carlo generators}
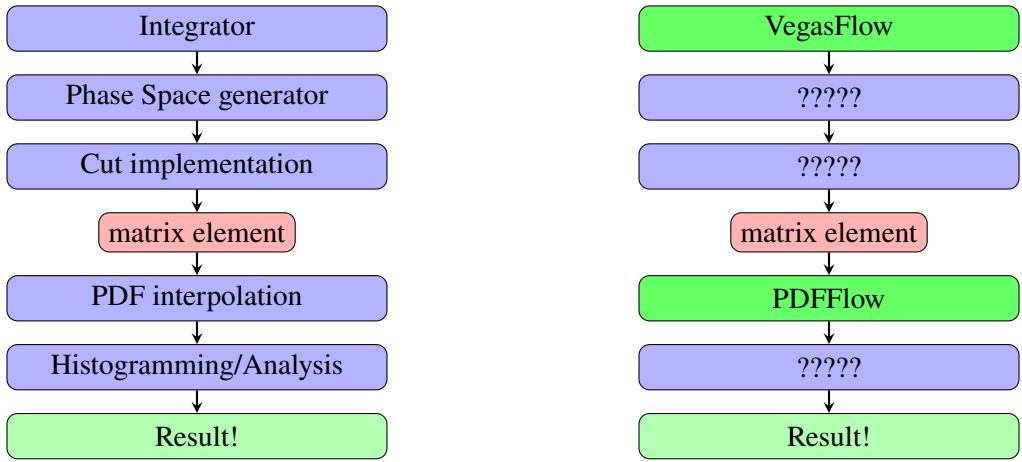
\begin{figure}[h]
    \center
    \begin{subfigure}{0.45\textwidth}
        \centering
        \begin{tikzpicture}[node distance = 1.0cm]
            \node[roundbox] (integrator) {Integrator};
            \node[roundbox, below=0.3cm of integrator] (ps) {Phase Space generator};
            \node[roundbox, below=0.3cm of ps] (cuts) {Cut implementation};
            \node[startstop, below=0.3cm of cuts] (me) {matrix element};
            \node[roundbox, below=0.3cm of me] (lhapdf) {PDF interpolation};
            \node[roundbox, below=0.3cm of lhapdf] (hist) {Histogramming/Analysis};
            \node[roundbox, fill=green!30, below=0.3cm of hist] (result) {Result!};

            \draw[arrow] (integrator) -- (ps);
            \draw[arrow] (ps) -- (cuts);
            \draw[arrow] (cuts) -- (me);
            \draw[arrow] (me) -- (lhapdf);
            \draw[arrow] (lhapdf) -- (hist);
            \draw[arrow] (hist) -- (result);
        \end{tikzpicture}
        \caption{Schematic view of the different ingredients that form a Monte Carlo generator}
        \label{fig:schemeCPU}
    \end{subfigure}
    \hfill
    \begin{subfigure}{0.45\textwidth}
        \centering
        \begin{tikzpicture}[node distance = 1.0cm]
            \node[roundbox, fill=green!60] (integrator) {VegasFlow};
            \node[roundbox, below=0.3cm of integrator] (ps) {?????};
            \node[roundbox, below=0.3cm of ps] (cuts) {?????};
            \node[startstop, below=0.3cm of cuts] (me) {matrix element};
            \node[roundbox, fill=green!60, below=0.3cm of me] (lhapdf) {PDFFlow};
            \node[roundbox, below=0.3cm of lhapdf] (hist) {?????};
            \node[roundbox, fill=green!30, below=0.3cm of hist] (result) {Result!};

            \draw[arrow] (integrator) -- (ps);
            \draw[arrow] (ps) -- (cuts);
            \draw[arrow] (cuts) -- (me);
            \draw[arrow] (me) -- (lhapdf);
            \draw[arrow] (lhapdf) -- (hist);
            \draw[arrow] (hist) -- (result);
        \end{tikzpicture}
        \caption{For GPU computation, most of the ingredients need to be coded from scratch.}
        \label{fig:schemeGPU}
    \end{subfigure}
    \caption{Schematic view of the ingredients necessary to write a new Monte Carlo generator for a new process. In this
    case the ideal situation is such we can only worry about the matrix element and use existing tool for the rest of
the program (left). For GPU computing, however, this cannot yet be the case, as only some of the tools exist (right).}
    \label{fig:scheme}
\end{figure}

Monte Carlo generators are at the core of HEP phenomenology, and they have been for many years.
As a result a plethora of tools, libraries and algorithms have been developed around these programs.
Some examples of this are the LHAPDF library~\cite{Buckley:2014ana}, the RAMBO~\cite{Kleiss:1985gy}
phase space generator or the Vegas algorithm mentioned above.
This brings us to the situation in Fig.~\ref{fig:schemeCPU} where, if one would like to write a new
program one can also take advantage of existing tools, greatly reducing the amount of effort necessary to produce results.

For GPU computations, however, the situation is much worse:
many (if not most) of the necessary tools need to be written from scratch.
Not only the ones we are interested in (in the example of Fig.~\ref{fig:schemeGPU}, the matrix element), but also
auxiliary elements such as the integrator or the PDF interpolation library.

With VegasFlow~\cite{Carrazza:2020rdn} and PDFFlow~\cite{Carrazza:2020qwu} we provide two of these tools
which we hope will kick-start a new era of truly performant HEP phenomenology where the new advances in hardware are
exploited to the fullest.

\section{VegasFlow}
GPU computing has become ubiquitous in the world of High Performance Computing (HPC).
These devices provide a way of enormously accelerating parallel computations reducing both computing time and power consumption.
With VegasFlow we aim to eliminate the technological gap between the HPC and HEP communities
by providing a library that can silently offload very complicated calculation to hardware accelerators.

A paramount example is the simulation of proton-proton collisions considered above,
with PDFFlow~\cite{Carrazza:2020qwu} we already offload the computation of the initial state to the GPU,
with VegasFlow~\cite{Carrazza:2020rdn} the numerical Monte Carlo is also offloaded to the GPU.
At this point only the actual matrix element, describing the physics of the interaction, and its phase space need to be
written by the user.

A detailed description of the code, together with examples, can be found in the documentation of the library
library~\cite{juan_cruz_martinez_2020_3691927}.

One of the challenges of HPC is to manage multidevice computation.
In this proceedings we will mention two of the use cases considered in VegasFlow:
running on multiple-GPUs in one single machine and running on a cluster.

\subsection{Multi-GPU}

The number of events that can be run in parallel on a GPU is directly related to the memory capacity of the device.
As a rule of thumb, running more events at once will reduce the latency associated with the communication between
the GPU and the CPU.
On the other hand, and depending on the specifics of the hardware, it might be preferable to reduce the number
of points per batch.
One reason to do so is to ensure that all GPUs (if there is more than one) receive a share of the data
so no device is idle at any point.

\begin{figure}
    \center
    \includegraphics[width=.8\textwidth]{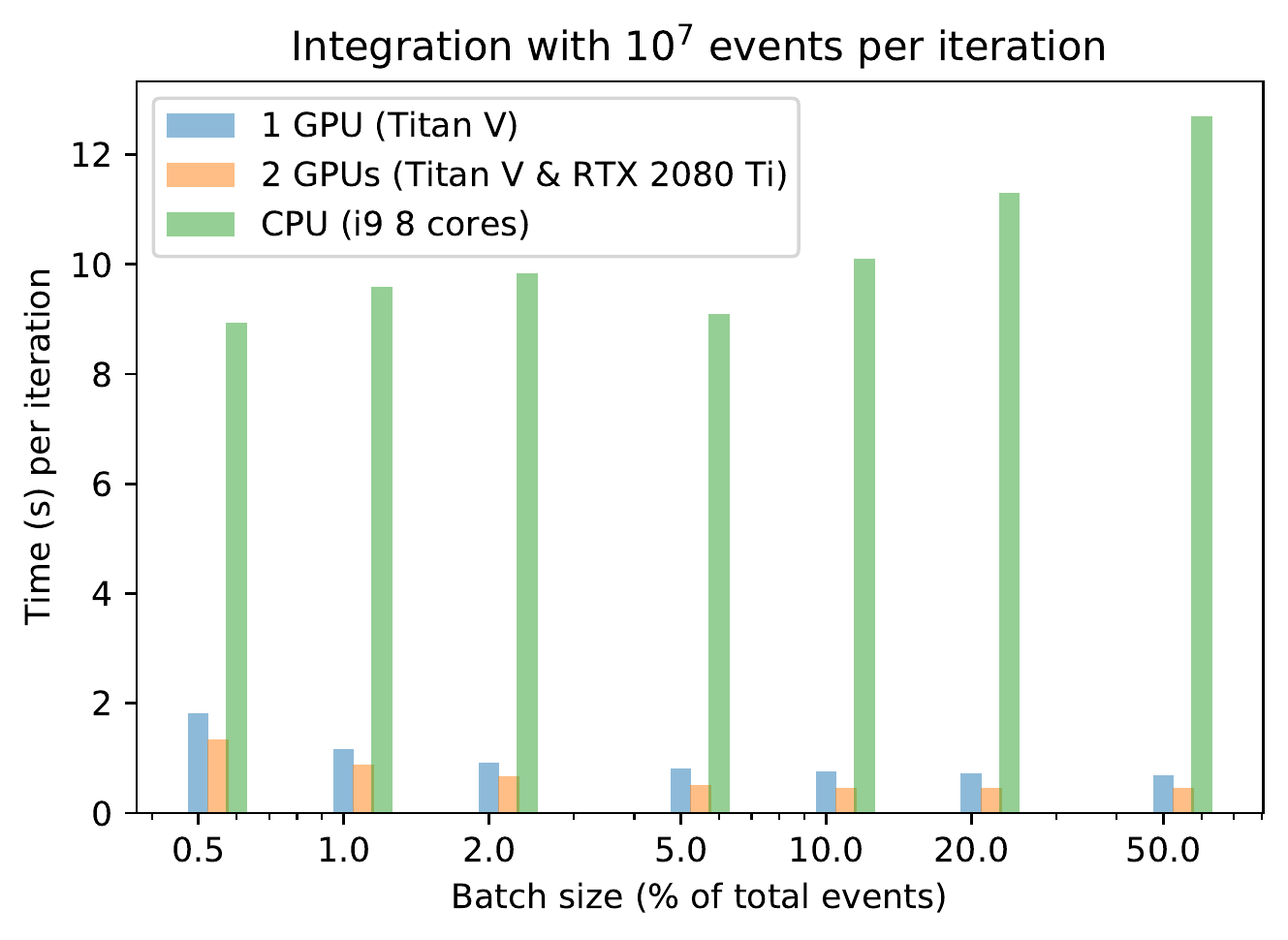}
    \caption{Example of the effect of running an integration on different devices with different batch sizes. Greater
        batch sizes can improve the performance of the GPU by minimizing I/O latency. For CPU computation, where this
    issue doesn't exist, a greater batch size actually hurts performance. The CPU used for this example is a Intel
    i9-9980XE. The GPUs are NVIDIA's Titan V (working alone in 1 GPU mode) and GeForce RTX 2080Ti. The integration corresponds to the single top example shown
in~\cite{Carrazza:2020rdn}.}
    \label{fig:eventlimit}
\end{figure}

In order to control the amount of points that VegasFlow will send to the GPU we provide the keyword
\texttt{events\_limit}.
This variable becomes especially relevant for multidevice computation.
It is clear from Fig.~\ref{fig:eventlimit} that there is no one-trick value as different devices will work
differently as the batch size changes.
We observe that the performance of the calculation on the GPU is impacted when the number of events per batch
is too small.
This can be easily understood, as for a batch size of 50\% of the events, there are only two transfers of data
from the CPU to the GPU, while for a batch size of 2\% there will be 50 data transferring operations.
We also observe that, once a certain threshold is reached, increasing the size of the batch does not impact the
performance.
This threshold depends on the hardware (how fast is the data transfer) and the calculation
(for how long will the GPU be computing) and it is the point at which the transfer time is offset by the calculation
itself.

Below we show an example of how to set the batch size using \texttt{events\_limit} in actual code with a small toy 
integrand:

\begin{minted}[autogobble]{python}
    >>> from VegasFlow import VegasFlow
    >>>
    >>> def complicated_integrand(xarr, **kwargs):
    >>>     return tf.reduce_sum(xarr, axis=1)
    >>>
    >>> n_dim = 10
    >>> n_events = int(1e6)
    >>> integrator = VegasFlow(n_dim, n_events, events_limit = int(1e5))
    >>> integrator.compile(complicated_integrand)
    >>> res = integrator.run_integration(n_iter = 5, log_time = True)
\end{minted}

Finally, one other interesting aspect on Fig.~\ref{fig:eventlimit} is that for the CPU the behaviour is opposite to that
of the GPU. In this case there is no transfer of data to be performed and so an increase of the batch size has a
negative impact as VegasFlow needs to allocate a bigger chunk of RAM.

\subsection{Cluster}
A common use case for any parallelizable Monte Carlo algorithm is cluster computation.
In the simplest scenario one would just send a separate instance of the program to each node of the cluster,
combining the results afterwards in a consistent way.

This, however, presents a challenge for adaptable Monte Carlo algorithm such as VegasFlow, as every iteration
informs the subsequent run.
We overcome this problem by implementing an interface to the Dask~\cite{dask} library.

The main advantage of implementing an interface to Dask is that it means automatic support for all Dask-supported backends.
The authors must warn, however, that it has not been possible to test it beyond the systems available to them,
so only the SLURM~\cite{slurm} system can be guaranteed to work.

Below we extend the previous example where a \texttt{SLURMCluster} instance from Dask is configured and passed to VegasFlow.
After calling the method \texttt{set\_distribute} with the reference to the chosen cluster, the method
\texttt{run\_integration} will send the job to the cluster instead of running them in the local computer.
An example on how VegasFlow scales is given in Table~\ref{table:dask}.

\begin{minted}[autogobble]{python}
    >>> from dask_jobqueue import SLURMCluster
    >>>
    >>> cluster = SLURMCluster(queue="<q>", project="<p>", cores=4, memory="2g")
    >>>
    >>> integrator.set_distribute(cluster)
    >>> res = integrator.run_integration(n_iter)
\end{minted}

\begin{table}[h]
    \centering
    \begin{tabular}{c c c c}
        \hline & 1 node & 2 nodes & 3 nodes \\ \hline
        Time per iteration (s) & 59.7s & 33.3s & 22.3s \\ \hline
    \end{tabular}
    \caption{How an integration scales in a SLURM cluster as a function of the number of nodes active. Using 4 cores per node.
        Note that time does not scale linearly with the number of nodes as a greater number of nodes also means a
        greater number of data transferring operations.}\label{table:dask}
\end{table}

\section{Conclusions}
We have demonstrated the performance gains that can be obtained from hardware accelerators (in a single node or in a
computing cluster) by using the VegasFlow library.
We show how the improvement is sensitive to the distribution strategy. Indeed, in a fashion similar to machine learning
techniques, finding the right batch size can make a notable difference on the performance.

\acknowledgments

The authors are supported by the European Research Council under the European Union's Horizon 2020 research and
innovation Programme (grant agreement number 740006) and by the UNIMI Linea 2A grant ``New
hardware for HEP''.

\bibliographystyle{../JHEP}
\bibliography{../blbl}

\end{document}